\documentclass[preprint,12pt,authoryear]{elsarticle}

\usepackage{amssymb}
\usepackage{textcomp}
\usepackage{inputenc}
\usepackage{longtable}
\usepackage{subcaption}
\usepackage{outlines}
\usepackage{xurl}
\usepackage[symbol]{footmisc}

\journal{arXiv}

\begin{document}

\begin{frontmatter}


\renewcommand{\thefootnote}{\fnsymbol{footnote}}
\title{Large Scale Analysis of Open MOOC Reviews to Support Learners' Course Selection\footnote[2]{Funding: this research did not receive any specific grant from funding agencies in the public, commercial, or not-for-profit sectors.}}


\author[inst1]{Manuel J. Gomez\corref{mycorrespondingauthor}}
\ead{manueljesus.gomezm@um.es}
\affiliation[inst1]{organization={Faculty of Computer Science},
            addressline={Calle Campus Universitario}, 
            postcode={30100}, 
            state={Murcia},
            country={Spain}}
          
\affiliation[inst2]{organization={SkillMapper},
            addressline={6 rue de Steinkerque}, 
            postcode={75018}, 
            state={Paris},
            country={France}}

\author[inst2]{Mario Calderón}
\ead{mario@skillmapper.com}
\author[inst2]{Victor Sánchez}
\ead{victor@skillmapper.com}
\author[inst1]{Félix J. García Clemente}
\ead{fgarcia@um.es}
\author[inst1]{José A. Ruipérez-Valiente}
\ead{jruiperez@um.es}

\cortext[mycorrespondingauthor]{Corresponding author}

\begin{abstract}
The recent pandemic has changed the way we see education. It is not surprising that children and college students are not the only ones using online education. Millions of adults have signed up for online classes and courses during last years, and MOOC providers, such as Coursera or edX, are reporting millions of new users signing up in their platforms. However, students do face some challenges when choosing courses. Though online review systems are standard among many verticals, no standardized or fully decentralized review systems exist in the MOOC ecosystem. In this vein, we believe that there is an opportunity to leverage available open MOOC reviews in order to build simpler and more transparent reviewing systems, allowing users to really identify the best courses out there. Specifically, in our research we analyze 2.4 million reviews (which is the largest MOOC reviews dataset used until now) from five different platforms in order to determine the following: (1) if the numeric ratings provide discriminant information to learners, (2) if NLP-driven sentiment analysis on textual reviews could provide valuable information to learners, (3) if we can leverage NLP-driven topic finding techniques to infer themes that could be important for learners, and (4) if we can use these models to effectively characterize MOOCs based on the open reviews. Results show that numeric ratings are clearly biased (63\% of them are 5-star ratings), and the topic modeling reveals some interesting topics related with course advertisements, the real applicability, or the difficulty of the different courses. We expect our study to shed some light on the area and promote a more transparent approach in online education reviews, which are becoming more and more popular as we enter the post-pandemic era.
\end{abstract}


\end{frontmatter}


\section{Introduction}
\label{sec:introduction}

Massive Open Online Courses (MOOCs) are a relatively recent educational phenomenon that has received a great deal of attention during the last decade \citep{pina2015moocs}. This attention is due to their potential to disrupt the traditional educational pathways trough ease of access and free or low-cost contents. MOOCs offer the potential to enable access to high-quality education to students, even in the most underserved regions of the world \citep{castillo2015moocs}. Furthermore, the recent COVID-19 pandemic has completely shifted the behavior of students around the world \citep{dhawan2020online}. Strict lockdowns and a growing fear of returning to classrooms were the perfect storm for the unparalleled acceleration in the adoption of MOOCs. Thus, it is not surprising that the largest providers of MOOCs such as Coursera or Udemy reported a +50\% growth in revenue in 2020 \citep{lohr2020remember}, the most significant year-over-year increase since the beginning of the MOOC era. Taking advantage of the momentum, ed-tech providers attempted to capitalize by widening their offer of courses. In fact, Class Central reported that, by the end of 2020, 16,300 MOOCs will be announced or launched, and in 2020 alone, around 2,800 courses were added \citep{classCentralReport}.

Though a sudden increase in the educational offer seems like an idyllic scenario as it provides a broader range of educational options, it is yet to be confirmed whether the aforementioned is net positive for the system as a whole. Some would argue that a supply-side where providers like Coursera alone shows around 415 Excel courses simply does not need one more Excel course. As a result, we might be reaching what some scholars call ``choice overload,'' as a higher number of options might decrease the motivation to choose or the satisfaction with the final preferred option \citep{scheibehenne2010can}. Arguably, this lack of satisfaction with the preferred option contributes to low completion rates in the MOOC space, which are reported to be around 10\% \citep{khalil2014moocs}. Moreover, \cite{hew2014students} reported that the quality of MOOC education and the MOOC business model are still unresolved issues. In addition, more than 55\% of the students who were part of a research conducted in LATAM reported spending days and weeks choosing a course \citep{skillmapper2022}; thus, shedding light on how time-consuming the process of finding and purchasing courses could be. A process that ultimately fails to deliver on the promise of democratizing education.

In addition to the trend above, students do face other challenges when choosing courses. Though online review systems are standard among many verticals, no standardized or fully decentralized review systems exist in the MOOC ecosystem. Whereas some MOOC providers rely on the standard 5-star rating system and free form feedback, such as Class Central \citep{classC}, other popular ones do not even offer the alternative to leave a single review. Moreover, many of these MOOCs only allow the user to leave a review once the course has been completed \citep{gamage2016star}. Thus, the opinions of around the 90\% of students who never finished the MOOC are missing, hence usually creating a very positive and skewed perspective. In this vein, NCA (research agency) found that around 18\% of students that do not finish a MOOC experienced teachers with low pedagogical qualities \citep{skillmapper2022}. Naturally, the restrictiveness of many of these review systems limits the voice of many students, thus suppressing the distribution of these kinds of relevant insights for MOOC purchase decision making. This transparency in the reviewing systems could allow new approaches to facilitate students' choices when comparing different courses.          

The inconsistent review systems, high search costs, and meager completion rates contribute to many MOOC providers’ poor experience. Current textual reviews are considered as additional information to the rating, to know how the course is, but it is not considered in the rating of the review itself and in personalized recommendations \citep{kapoor2020movie}. We believe that there is an opportunity to leverage available open MOOC reviews, in order to build simpler and more transparent reviewing systems, allowing users to really identify the best courses out there. In fact, Natural Language Processing (NLP) provides us a set of techniques and approaches to analyze textual information (e.g., sentiment analysis, topic modeling, text classification) that could be used to add value to the reviews and provide useful insights for learners. Ultimately, there is an opportunity to provide a framework that could benefit the overall MOOC industry. Thus, enable more transparency, better discoverability, and increased satisfaction with choice, also helping to greatly improve the professional development of our society to current industry needs. As a by-product, these enhancements could also lead to improved overall satisfaction with MOOC providers. In our work, we aimed to collect the largest dataset of MOOC reviews from five different platforms via web scraping, in order to analyze around 2.4 million reviews and to transform these big data into useful insights and actionable information for learners and other stakeholders, such as industry and educators. Specifically, we state four research questions (RQs):

\begin{outline}
    \1 \textbf{RQ1.} Do the crowdsourced numeric ratings of the courses provide discriminant information to prospective learners?
    \1 \textbf{RQ2.} Does NLP-driven sentiment analysis on the textual reviews provide valuable information to prospective learners?
    \1 \textbf{RQ3.} Can we leverage NLP-driven topic analysis techniques to find themes that can be important for prospective learners?
        \2 \textbf{RQ3.1.} Using words describing the course in a qualitative way.
        \2 \textbf{RQ3.2.} Using words describing the courses' content.
    \1 \textbf{RQ4.} Can we use these models to effectively characterize MOOCs based on the open reviews?
\end{outline}

The rest of the paper is organized as follows. Section \ref{sec:relWork} reviews background literature on MOOCs, previous course selection approaches, and previous analyses using MOOCs reviews. Section \ref{sec:methods} describes the complete methodology, and Section \ref{sec:results} shows the results of our analysis, answering each one of the RQs. Then, we finalize the paper with discussion in Section \ref{sec:discussion} and conclusions and future work in Section \ref{sec:conclussion}. 

\section{Related Work}
\label{sec:relWork}

Technological advances, particularly in the Web 2.0, have brought a major transformation in the delivery of education, and one of the most recent innovations in e-learning is the development, roll out and uptake of MOOCs \citep{terras2015massive}. MOOCs represent the next stage in the evolution of open educational resources \citep{al2018model}. In addition, they have the advantage of being available for all and open to an unlimited number of students \citep{al2019massive}. The increase in numbers of MOOCs has been dramatic in recent years, and their scope, available in every sector of education, makes them very strong and being considered to be a new form of virtual technology enhanced learning environments \citep{pina2015moocs}. In 2013, \cite{billington2013moocs} stated that some experts considered MOOCs as a bubble that would burst soon. However, by the end of 2020, more than 180 million students have registered in one or more of the 16,300 courses offered by over 950 universities worldwide \citep{classCentralReport}, indicating that MOOCs continued to capture the attention of many institutions and the public worldwide since the ``year of the MOOC'' in 2012 \citep{pappano2012year, lan2020examining}. Moreover, they have overcome all the challenges of the COVID-19 pandemic and have been instrumental in providing courses to the learners as they do not have physical boundaries \citep{purkayastha2021unstoppable}.

Although MOOCs have a great potential for educating large numbers of students in a wide variety of settings, previous research has also identified some problematic issues with this area. One of the major problems is the high dropout rate for learners: a  small percentage (generally around 10\%) of the large numbers of participants enrolling in MOOCs manage to complete the course, which is a poor result compared to traditional courses \citep{liyanagunawardena2014dropout}. \cite{bezerra2017review} conducted a review on the different reasons causing this high dropout rates, identifying 24 different reasons, such as the lack of prior knowledge, the difference between the expected and the real level of the course, or the quality of materials. The second major area of concern relates to the financial model used in these courses. While there is a great potential for attracting large numbers of students, universities still need to determine how to generate sufficient income from the delivery of these free or low-priced courses \citep{milheim2013massive}. We see another example in \citep{jobe2014moocs}, where authors reported challenges associated with recognition, validation, and accreditation of learning, even though this accreditation is obviously a difficult goal to achieve. Of course, another area of major concern is academic integrity, including verifying that a student who is registered for a course is the same person who completed the tasks, and the potential for widespread plagiarism or homework assignments. Finally, another common problem that usually is not discussed in the literature is the course selection. Given the large numbers of courses (most of them with very high ratings) and information out there, learners are easily disoriented by the information overload \citep{zhang2018mcrs}, and they usually find difficulties choosing which is the appropriate course for them (taking several days to decide); this course selection difficulty is also related to the high dropout rate as they may not be choosing the best course to fit their preferences. Some of the reasons of the high dropout rate reported previously could be related with the overload and misunderstandings that learners suffer, since they do not usually know the real characteristics of the course that they are choosing: in a survey conducted by \cite{gutl2014attrition}, 8.96\% of the students reported that the course was too difficult, meanwhile 7.46\% emphasized that the course was not challenging. Moreover, 14.93\% of students also highlighted that the learning environment was not personalized, and another 6.72\% indicated that the courses were poorly taught. The selection of courses is a crucial decision for learners, and the use of existing reviews from other learners to reveal the real courses' characteristics could be a interesting solution to make this choice easier and fruitful. 

Learners are exposed to various challenges with this excess of learning resources and information: which provider they have to choose to search for a specific MOOC? Who is the best provider? How to search a specific MOOC? Are the available MOOC reviews reliable? \citep{ouertani2017moocs}. Usually, in literature we find different approaches trying to perform an efficient course selection: collaborative filtering (according to users in same interests on same courses), content-based filtering (using courses features from courses that the user likes to recommend similar items), knowledge-based approaches, or hybrid recommender approaches, among others \citep{al2016automated}. \cite{ouertani2017moocs} presented the idea of a system that could satisfy learners' needs when searching for suitable courses among different providers, using previous experiences of its users. Moreover, \cite{al2016automated} developed a recommendation system that employed an association rules algorithm to recommend university elective courses to a target student based on what other similar learners have taken. \cite{aher2012prediction} presented an approach combining of k-means clustering and Apriori association rule algorithms are useful in recommending courses to students. We found many other studies that used these type of approaches to improve course selection \citep{aher2014aa, huang2019score, vanitha2019collaborative}. In fact, in their review, \cite{khalid2020recommender} reported that most of the work on the implementation of the recommender system for courses used collaborative and content-based filtering. In our research, we decided to use a novel approach that aims to leverage existing open MOOC reviews using NLP to gather information that could be useful for learners' course selection.

Moreover, as we do in our case study, previous researchers have explored the potential of using open MOOC reviews and NLP. For example, authors in \citep{deng2021key} reported that it is critical that student reviews are systematically analyzed and interpreted in conjunction with student ratings data. In their work, they presented a study using \textit{Leximancer}, a data-mining tool which extracted the key concepts from the data collection, based on the frequency of occurrence of concepts and co-occurrence use. This study highlighted some interesting key terms inferred, such as ``videos,'' ``engaging,'' ``understanding,'' ``easy,'' or ``problems.'' Furthermore, \cite{wen2014sentiment} applied sentiment analysis toward the curriculum and key course tools from forum posts, finding that the emotion of students was correlated with the number of students dropping out of the course each day. Moreover, \cite{chen2020moocs} employed an innovative structural topic modeling technique to analyze 1,920 reviews of 339 courses regarding computer science to understand what primary concerns the learners had. They found that 64.2\% of the reviews in their data collection were 5-star reviews, while 16.7\% were 4-star reviews, revealing clearly biased rating values. Regarding structural topic modeling, they found topics such as ``Course levels,'' ``Teaching style,'' or ``Course content.'' Furthermore, \cite{chang2016developing} presented a novel approach that used the courses' videos to locate ``hot'' video segments (i.e., the parts that students reviewed more) to generate their subtitles and extract the main keywords, being  useful for the teacher to better understand which parts of the contents of each topic are most difficult for the learners. In our work, we go beyond the state of the art by presenting an approach that uses the largest MOOC reviews dataset to date in the literature to build NLP models (specifically topic finding and sentiment analysis) oriented to provide help and support learners' course selection easily and quickly, and not only extracting terms or revealing important topics within a word collection. While the literature usually considers extracting a set of key terms from the entire reviews dataset, our topic modeling approach considers two different aspects of the reviews: on the one side, we consider words in reviews describing the course in a qualitative way; on the other side, we consider words in reviews describing the courses' content. This approach aims to gather insights from two different points of view, also reducing the bias produced by other types of words. 

\section{Methodology}
\label{sec:methods}



We conducted our analysis in six different stages: 1) MOOC review providers, 2) Data scraping, 3) Data cleansing and wrangling, 4) Data collection, 5) Data pre-processing, and 6) Data modeling. We can see the entire methodology process represented in Figure \ref{fig:methodFig}. Next, we explain each stage in detail:

\begin{figure}[!ht]
\includegraphics[width=\textwidth]{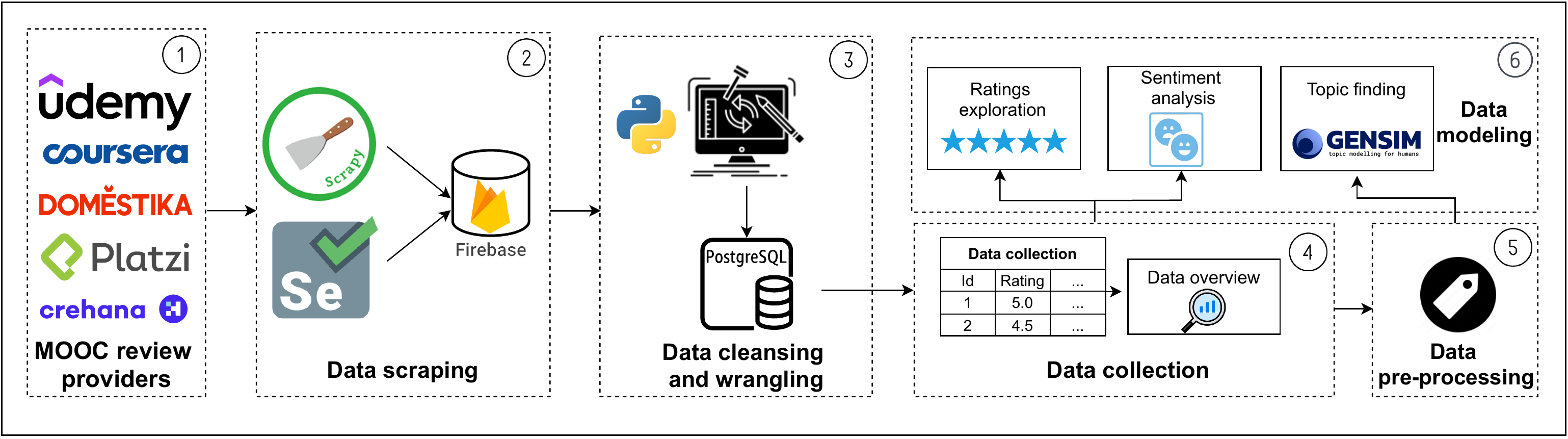}
\centering
\caption{Methodology diagram.}
\label{fig:methodFig}
\end{figure}

\subsection{Selection of MOOC review providers}

Initially, we considered five different MOOC review providers for our analysis: Udemy, Coursera, Domestika, Platzi, and Crehana. Next, we make a brief introduction to each platform:

\begin{itemize}
    \item \textbf{Udemy} is a for-profit MOOC provider aimed at professional adults and students. Founded in May 2010 by Eren Bali, Gagan Biyani, and Oktay Caglar, offers more than 183,000 courses taught by 65,000 different instructors and with 75 different languages available. It offers interesting features such as course quality checklists, a teaching hub, and training videos \citep{udemyRef}.
    \item \textbf{Coursera} is an American MOOC provider founded in 2012 by Andrew Ng and Daphne Koller. In 2021, 150 universities offered more than 4,000 courses through Coursera, reaching more than 190 different countries. The platform is updated every year with new features such as personalized browsing, learner skill tracking, or cloud imports \citep{courseraBlog, courseraAbout}.
    \item \textbf{Domestika} is an online learning platform for creatives, with paid courses on software, crafts, illustration, and more. It was first started more than a decade ago in Spain, but has since moved its headquarters to San Francisco. Although there are a few courses about Python and web programming like in other platforms, what makes the platform unique is its unwavering focus on creatives, and its major feature is the exceptionally high course quality \citep{domestikaAbout}.
    \item \textbf{Platzi} is a MOOC provider, founded in Colombia by Freddy Vega and Christian Van Der Henst in 2014. It offers more than 700 open courses within different areas, such as data science, videogames, or blockchain, among many others \citep{plaztiRef}.
    \item \textbf{Crehana} is an online education provider for today's student: digital and creative. Over 1,000,000 students around the world have used the platform, which offers subjects such as illustration, design, business and more. Its main goal is to bring education closer to all types of students regardless of geographical location and income level \citep{crehanaRef}.
\end{itemize}

\subsection{Data scraping}

First of all, we needed to retrieve the structure for each of the websites. We used two different approaches, depending on how the websites were structured. In our experience:

\begin{itemize}
    \item We used \textit{Scrapy} to work with websites that consider static pages. \textit{Scrapy} is a web framework built upon years of experience in extracting massive amounts of data in a robust and efficient manner \citep{kouzis2016learning}.
    \item We used \textit{Selenium}, a portable open source software \citep{bruns2009web}, to work with websites that are more dynamic, consider infinite scrolls or that need to click buttons to get more information.
\end{itemize}

Using these tools, we started scraping the existing reviews in each one of the platforms. We worked with the metadata divided into two different levels:

\begin{itemize}
    \item \textbf{Review level}. For each review, we collected the following information: URL, review, rating of the review (from one to five), the platform, the username, the date of the review, and the identifier of the related course.

    \item \textbf{Course level}. For each course, we collected the following information: the course identifier, URL, title, platform related with the course, the content category of the course, and the name of the teacher.
\end{itemize}

The first step was to download all the data crawled and store it directly in a database. Because there is no structure defined and we only need to process the data once, we used Firestore, a non-relational database from Google Cloud. Since Firestore provides a simple and cheap solution to store data in a  non-relational format, thus meeting our requirements. We used a dictionary structure to store the data in Firestore, adapted to each one of the educational platforms.

\subsection{Data cleansing and wrangling}

After the data is saved in the non-relational database, we proceeded to apply a step wise approach to clean and process the data, ensuring that the fields are properly registered. This part of the process is one of the most essential steps because it considers a set of rules that we use to process the metadata and give a proper structure to it. These series of rules mostly leverage a set of regular expressions and variable conversions. These functions allow us to remove elements that do not add value, such as commas, special characters, and normalize types of variables when necessary. As a result, we can obtain clean dataset that will help to optimize our analysis later. This section is ran in Python and consist in: (1) read all the data that we previously stored in Firestore, (2) apply a particular set of rules for each of the rows, and (3) save the processed value in an object.

Since we needed to identify the language of each review, and this information is not available in the original metadata, we identified it using the raw text of each review. Initially, we considered using the audio feature of the course to identify the language, but we can not rely on this feature to ensure that all the reviews are in the same language as the course, so we finally decided to work with \textit{Fasttext} library \citep{bhattacharjee2018fasttext} to infer the language of each review. 

The last step was to store the data in a PostgreSQL database. This one is a relational database that will help us to understand the data collected running some analytics easily. The objects that we built are inserted in each of the two tables that we consider for this project:
\begin{itemize}
    \item \textbf{final\_courses\_info}: contains all the metadata at course level.
    \item \textbf{final\_reviews\_info}: contains all the metadata at review level.
\end{itemize}
    
\subsection{Data collection}

We initially considered all the data scraped from the five different platforms to collect the metadata. However, since for this study we decided to focus only on English reviews, the data collection was distributed as follows:

\begin{itemize}
    \item \textit{Udemy} platform: It has 95,143 courses and 71,518 of them have reviews, meaning that 24.8\% of them do not have any review.
    \item \textit{Coursera} platform: It considers 4,154 courses and 2,932 of them have reviews, so 29.4\% of courses do not have any review.
    \item \textit{Domestika} platform: It has 558 courses in English and 541 of them have reviews, meaning that 3\% of the courses do not have any review.
    \item \textit{Platzi} platform: It has 451 courses in English and 450 of them have reviews, meaning that 0.002\% of the courses do not have any review.
    \item \textit{Crehana} platform: It has 1 course with reviews (100\%).
\end{itemize}

\begin{figure}[!ht]
\centering
    \begin{subfigure}{.51\textwidth}
      \centering
      \includegraphics[width=\linewidth]{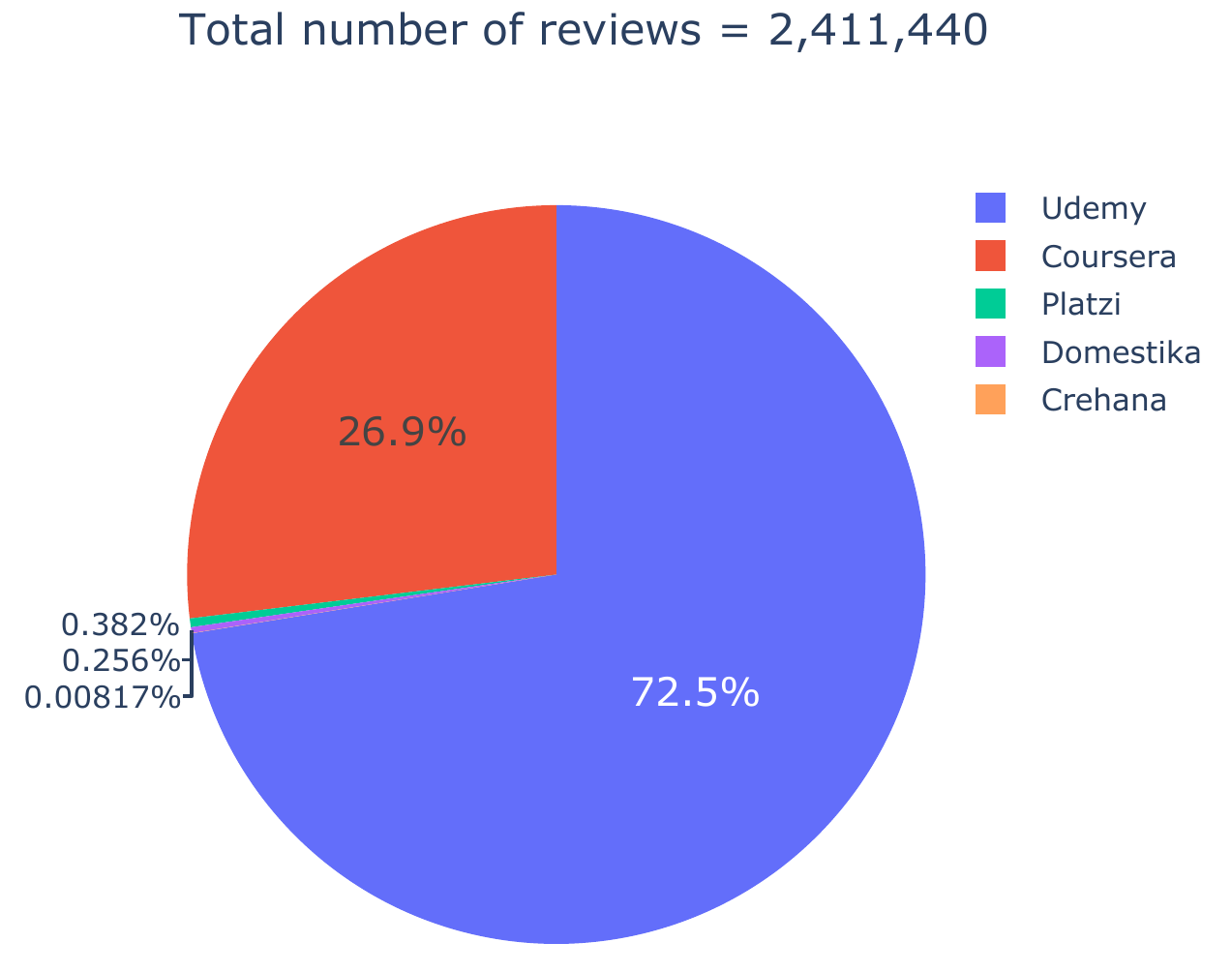}  
      \caption{Distribution of reviews by course provider.}
      \label{subfig:distrProv}
    \end{subfigure}
    \begin{subfigure}{.48\textwidth}
      \centering
      \includegraphics[width=\linewidth]{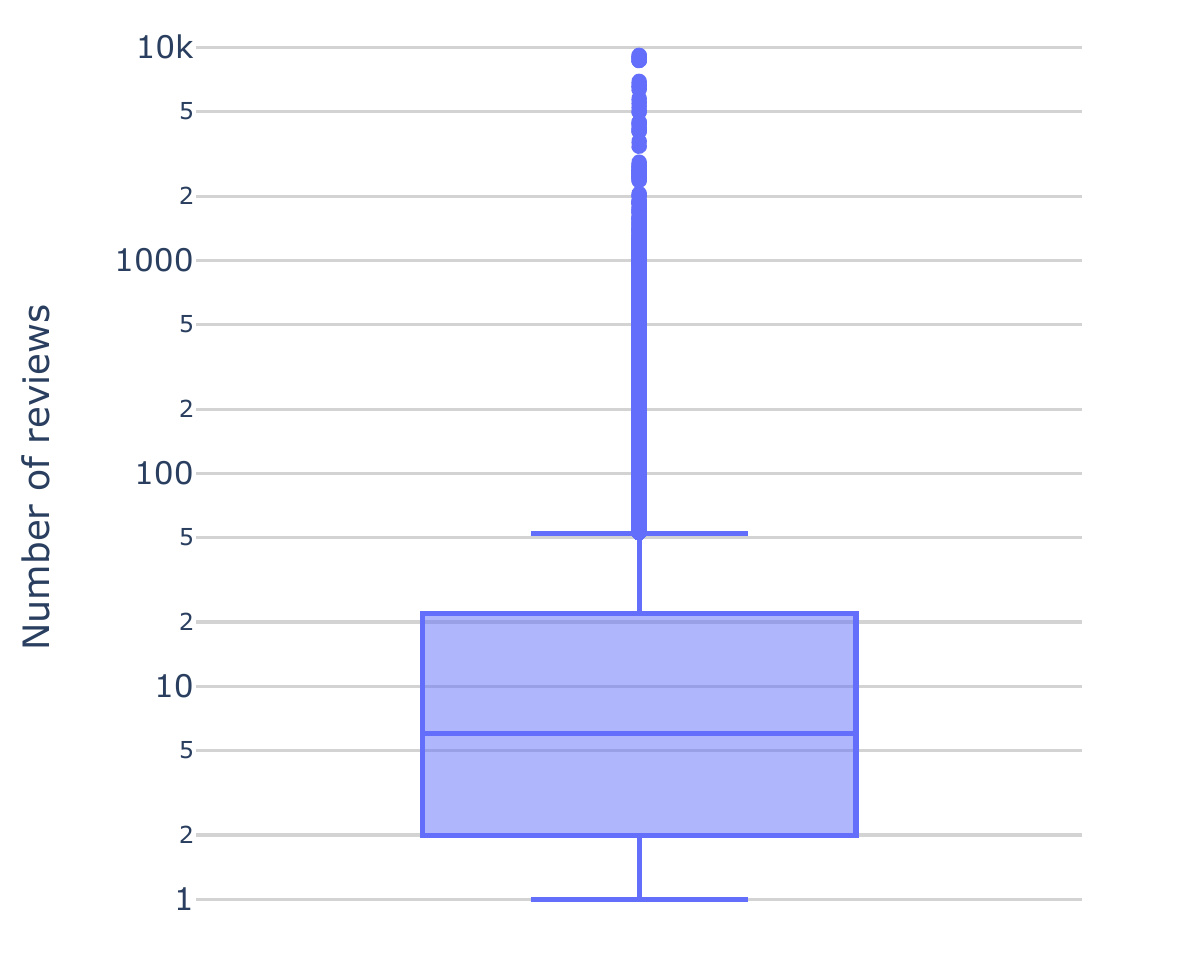}  
      \caption{Distribution of number of reviews per course.}
      \label{subfig:distrCours}
    \end{subfigure}
    \caption{Data collection overview.}
    \label{fig:dataColFig}
    \end{figure}

We established a threshold of one minimum review for each course to be considered for the analysis, meaning that 25\% of the courses were removed from our work. In our data collection, some of the reviews are concentrated on some courses. This can be related to several factors, such as important differences in popularity of each one of the courses, marketing efforts from the platforms, or hot topics. For example, a ``Roman history'' MOOC will attract a very different population of learners in quantity and background than a ``Python'' course. The distribution of reviews by the different providers can be seen in Figure \ref{subfig:distrProv}. Furthermore, in Figure \ref{subfig:distrCours} we can see the distribution of the number of reviews per course. The maximum number of reviews in a single one is 9,149, the minimum is one review, and the median is six reviews per course.

\subsection{Data pre-processing}

When trying to leverage NLP-driven topic finding in such a large set of reviews, we found that a large proportion of the words do not provide useful information to learners, such as ``lot,'' ``give,'' or ``thing,'' among many others. We wanted to provide feedback to learners that could become useful insights to select courses, thus we collected the most frequent words in the collection (words that appear more than 500 times in the entire collection), and we manually classified every word within two different categories:

\begin{itemize}
    \item \texttt{QualitativeDescription}: if the word is related to a qualitative description of the course (e.g., easy, clear, practical). This category includes 357 different words.
    \item \texttt{Content}: if the word is related to the content of the course itself (e.g., machine, yoga, cooking). This category includes 759 different words.
\end{itemize}

This categorization will help us to create the two different topic finding models in our research. Every word not matching any of those two categories will be excluded from our later analysis. Once we classified those words, the next step in the pre-processing is to clean each textual review. To do that, we keep only the review's main words removing, for example, unnecessary URLs, numbers, or additional space characters. Moreover, to apply NLP techniques afterward, we need to define a set of ``stop words'' (i.e., words that will not be considered in the text analysis). Some examples of stop words are ``would,'' ``be,'' or ``however,'' which are common words that appear in every document but do not provide helpful information to the analysis. 

From now, we can start treating each review as a ``document,'' since most of the cleaning process has ended. Once the full review collection cleaned, we lemmatize every document using \textit{pywsd} library. The lemmatization is the process of converting a word to its base form, and its implementation in \textit{pywsd} works as follows: 

\begin{enumerate}
    \item It tokenizes the string, dividing it into a set of tokens (words).
    \item It uses a Part-Of-Speech (POS) tagger to map each word to a POS tag (adverb, noun, adjective, etcetera).
    \item It calls the lemmatizer with the token and the POS tag to get the base form of the word.
\end{enumerate}

\subsection{Data modeling}

This subsection is divided into four separate parts, each one explaining the analysis to address each RQ:

\subsubsection{Ratings exploration}

To discover how was the distribution of ratings within the data collection, we checked two different approaches: in the first approach, we simply count every single numeric rating to see which are the most common ones; in the second approach, we calculated the mean rating of each one of the courses in order to see the distribution of ratings within the different courses separately.

\subsubsection{Sentiment analysis}
\label{sec:sentimentAnMeth}

The main aim of sentiment analysis (or opinion mining) is to discover emotion, opinion, subjectivity and attitude from a natural text. There are many different techniques that are frequently applied to natural text to determine the sentiment such as feature extraction, emoticon study, tokenization etc. During sentiment analysis, usually positive and negative words are extracted from the text and are assigned a score from the dictionary of words \citep{ahuja2017clustering}. In our work, to perform sentiment analysis, we explored the use of three different libraries in \textit{Python}, namely \textit{TextBlob}, \textit{VADER}, and \textit{Flair}:

\begin{itemize}
    \item \textit{TextBlob} is a Python (2 and 3) library for processing textual data. It provides a simple API for diving into common NLP tasks such as part-of-speech tagging, noun phrase extraction, sentiment analysis, classification, translation, and more \citep{loria2018textblob}. In his sentiment analysis implementation, given a sentence, it returns a score within the range [-1.0, 1.0], depending on how positive or how negative that sentence is.
    \item \textit{VADER}: Valence Aware Dictionary for Sentiment Reasoning (VADER) is a lexicon and rule-based sentiment analysis tool that is specifically attuned to the sentiments expressed in social media. It is an entirely free open-source tool, and it also takes into consideration word order and degree modifiers \citep{elbagir2019twitter}. Given its implementation, we used the ``compound'' value that the library provides, which is a number within the range [-1.0, 1.0], depending on how positive or how negative that sentence is.
    \item \textit{Flair} is a framework which aims to present a simple, unified interface for conceptually very different types of word and document embeddings, hiding all embedding-specific engineering complexity and allowing researchers to ``mix and match" various embeddings with little effort. It provides a pre-trained sentiment analysis model, which returns a tag (``Positive'' or ``Negative'') depending on the sentiment of the text provided.
\end{itemize}

\subsubsection{Topic finding}
\label{sec:topicFin}

To discover which are the main topics within the review collection, we applied Latent Dirichlet Allocation (LDA) topic modeling to the data provided. Specifically, we use \textit{gensim} library \citep{saxton2018gentle} and its \textit{ldaMallet} model \citep{graham2012getting}. Generally in LDA, each document can be described by a distribution of topics, and each topic can be described by a distribution of words; \textit{ldaMallet} uses an optimized Gibbs sampling algorithm for LDA \citep{yao2009efficient}.

There are multiple metrics for evaluating the optimal number of topics. Recent studies have shown that the classic predictive likelihood metric (or equivalently, perplexity) and human judgment are often not correlated, and even sometimes slightly anti-correlated \citep{o2015analysis}. This has led to many studies that have focused upon the development of topic coherence measures. To determine the optimal number of topics, we used two of these coherence measures: \textit{$C_v$} and \textit{$C_{umass}$} \citep{roder2015exploring,coherences}:

\begin{itemize}
    \item \textbf{\textit{$C_v$}} measure is based on a sliding window, one-set segmentation of the top words and an indirect confirmation measure that uses normalized pointwise mutual information (NPMI) and the cosine similarity.
    \item \textbf{\textit{$C_{umass}$}} is based on document co-occurrence counts, a one-preceding segmentation and a logarithmic conditional probability as confirmation measure.
\end{itemize}

Based on these two coherence measures, we chose the number of topics of the two final models calculated. The first of our models aimed to identify topics that described the course in a qualitative way (e.g., easy, clear, detailed, hard), thus we selected, for each review, only words that were classified as \texttt{QualitativeDescription}, removing the rest of the review to avoid bias produced by other type of words. Then, the second of our models aimed to identify topics that described the content of the courses, thus we selected only words that were previously classified as \texttt{Content}, removing the remaining ones.

Then, based on the topics discovered by the topic finding algorithm, we calculated the proportion of each topic across the entire corpus. To do that, we evaluated each review to get its topics associated (note that, in LDA, each document can be assigned to several topics with a certain weight). We calculated the proportion of each topic as follows:

\begin{equation}
\label{eq:propTopic}
Proportion\_topic_j = \frac{\sum_{i=1}^{N} weight\_topic_{ij}}{N} * 100
\end{equation}

Therefore, the proportion of topic $j$ would be the summation of each weight assigned to the topic $j$ in each document from $i$ to $N$, divided by the number of documents in the corpus ($N$). 

To provide a more complete analysis, we wanted to see the relationship between the sentiment analysis and the qualitative topic model (e.g. see if a course with a bad mean rating also is related to more negative topics). That way, we calculated the distribution of topics in each course separately (the distribution in each course will be the sum of proportions of each review separately), and compared them to the positive or negative values given by the sentiment classifier. A given course will be labeled as \texttt{Positive} if the majority of its reviews are labeled as \texttt{Positive}, and \texttt{Negative} if the majority of its reviews are labeled as \texttt{Negative}.

\section{Results}
\label{sec:results}

\subsection{RQ1. Do the crowdsourced numeric ratings of the courses provide discriminant information to prospective learners?}

\begin{figure}[!ht]
\centering
    \begin{subfigure}{.45\textwidth}
      \centering
      \includegraphics[width=\linewidth]{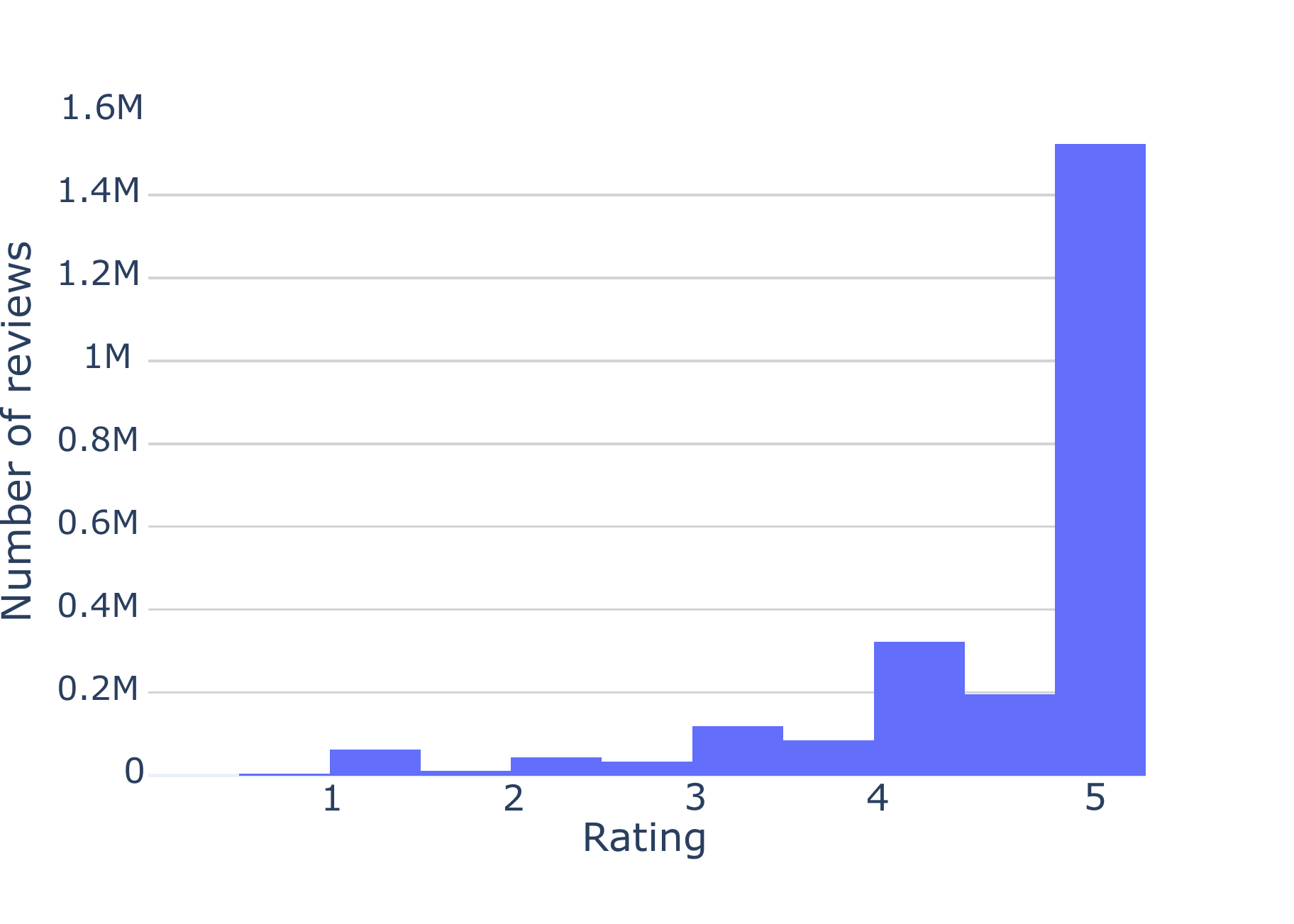}  
      \caption{Numeric ratings distribution.}
      \label{subfig:rq1Raw}
    \end{subfigure}
    \begin{subfigure}{.54\textwidth}
      \centering
      \includegraphics[width=\linewidth]{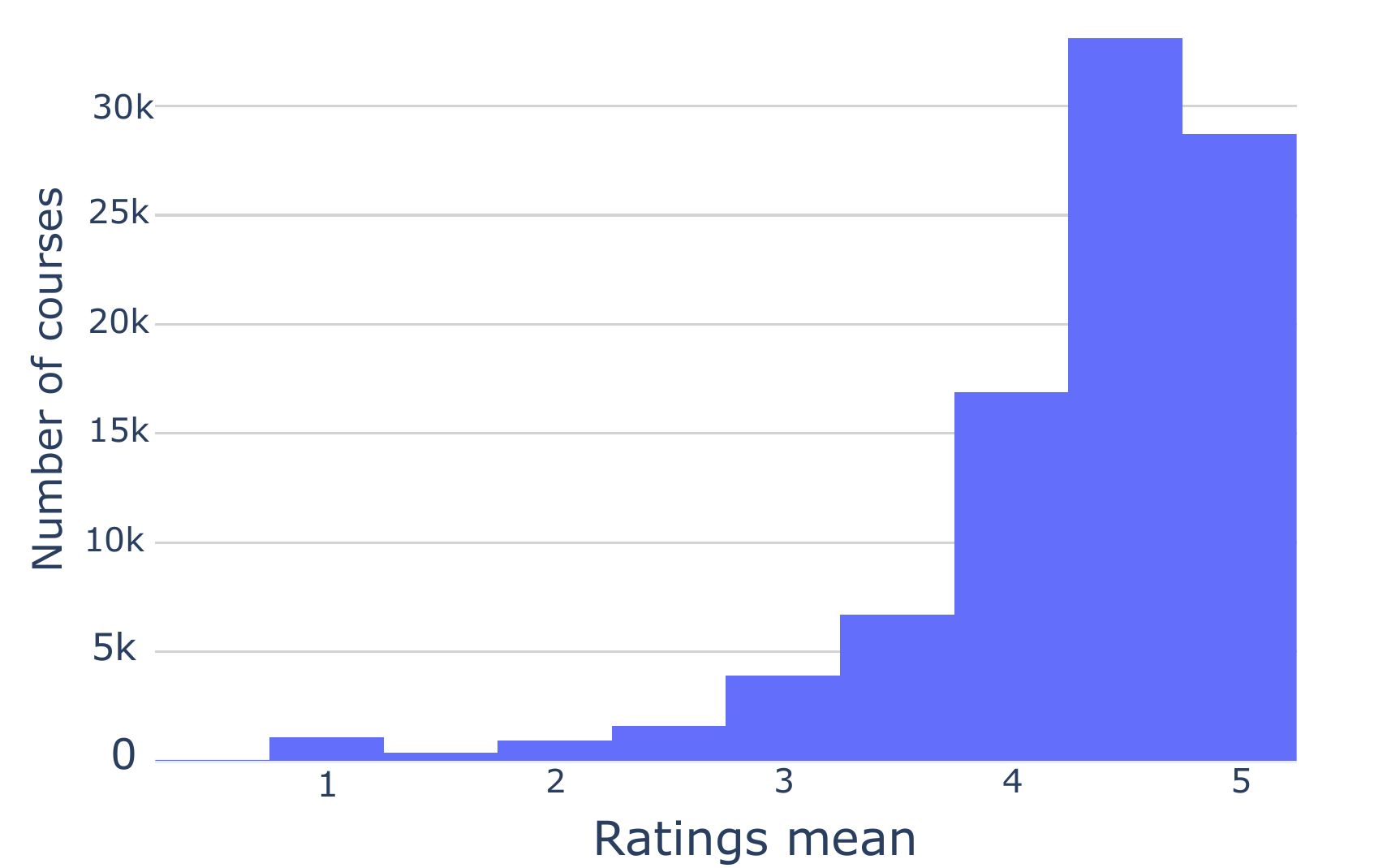}  
      \caption{Distribution of courses based on their ratings mean.}
      \label{subfig:rq1Mean}
    \end{subfigure}
    \caption{Raw ratings and course rating mean distribution.}
    \label{fig:rq1Fig}
    \end{figure}

In Figure \ref{subfig:rq1Raw} we can see the distribution of numeric ratings available in our collection. About 1.52 million reviews (63\% of the total) are 5-star ratings, and also about 519,000 reviews (21.5\% of the total) are 4 and 4.5 star ratings, thus we have a clearly biased data, where only 17.5\% of the ratings are 3.5-star ratings or worse. Moreover, in Figure \ref{subfig:rq1Mean} we see that, out of 93,678 total courses, 48,640 (52\%) have a mean rating between 4.5 and 5 stars, and 34,770 (37\%) courses have a mean higher of equal than 3.5 and below 4.5, highlighted again the systematic biases existing in open MOOC reviews. Therefore, this motivates alternative models to facilitate course selection to learners'.

\subsection{RQ2. Does NLP-driven sentiment analysis on the textual reviews provide valuable information to prospective learners?}

As we stated in Section \ref{sec:sentimentAnMeth}, we performed sentiment analysis on textual reviews based on three different libraries: \textit{TextBlob}, \textit{VADER}, and \textit{Flair}. Since \textit{TextBlob} and \textit{VADER} provide numeric values, we can compare both libraries directly. After applying the pre-trained sentiment models to every review in our corpus, we obtained a mean compound value of 0.54 (SD = 0.37) using \textit{VADER}, and a mean compound value of 0.37 (SD = 0.30) using \textit{TextBlob}. Then, we aggregated the sentiment values by course. On the one hand, we find that 67,019 courses (71.5\%) have a mean compound value above 0.4 (high positive mean) using \textit{VADER}. On the other hand, we find that 29,777 courses (31.8\%) have a mean compound value above 0.4 using \textit{TextBlob}.

We also calculated the Pearson's correlation coefficient and the Spearman's rank correlation coefficient between the compound sentiment value and the numeric ratings for both libraries. VADER library got a Pearson's coefficient value of 0.49, and a Spearman's coefficient value of 0.35. In the case of \textit{TextBlob}, the Pearson's coefficient was 0.38, while the Spearman's coefficient was 0.31. Although all these coefficients show a positive correlation between the two variables, VADER library obtained a higher value, indicating a moderate level of correlation since it is below 0.5.

\begin{figure}[!ht]
\centering
    \begin{subfigure}{.49\textwidth}
      \centering
      \includegraphics[width=\linewidth]{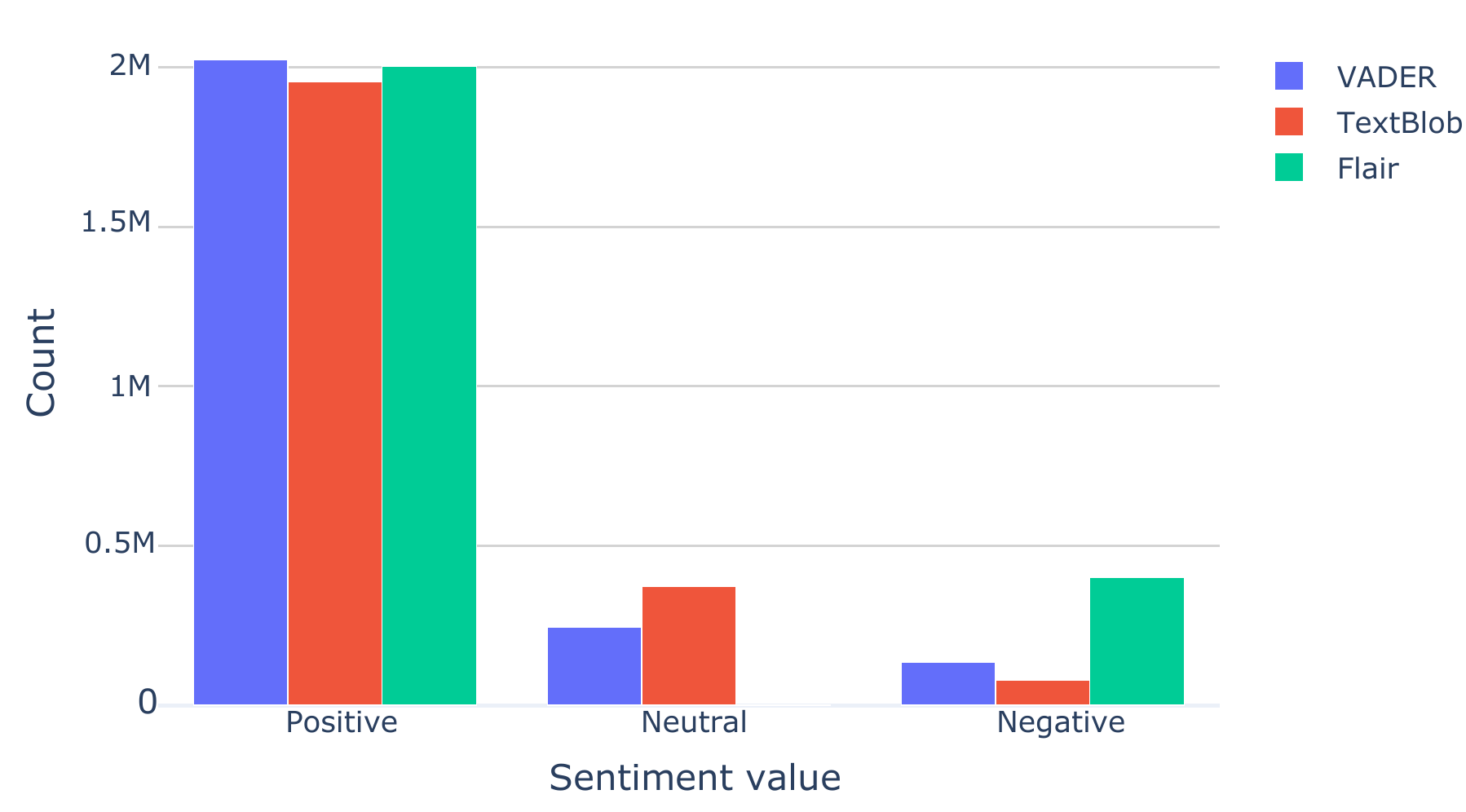}  
      \caption{Categorized sentiment analysis comparison between libraries.}
      \label{subfig:sentimentComp}
    \end{subfigure}
    \begin{subfigure}{.49\textwidth}
      \centering
      \includegraphics[width=\linewidth]{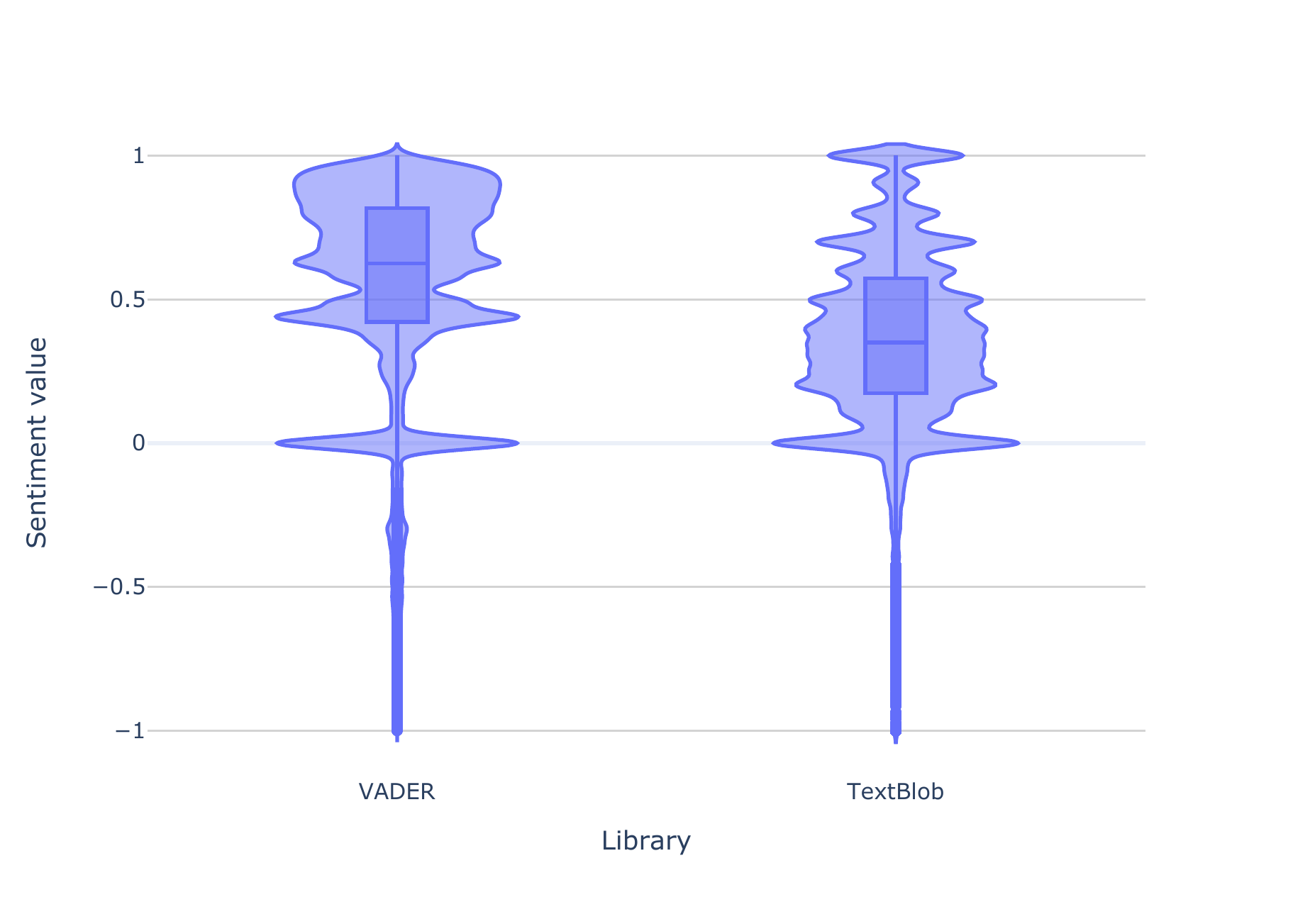}  
      \caption{Comparison of sentiment values between VADER and TextBlob library.}
      \label{subfig:boxplotSent}
    \end{subfigure}
    \caption{Sentiment results.}
    \label{fig:sentRes}
    \end{figure}

In the case of \textit{Flair}, it provides a label indicating if the sentence is \texttt{Positive} or \texttt{Negative}. To compare the three libraries, we established a threshold to classify \textit{TextBlob} and \textit{VADER} compound values into one of the three (positive, negative and neutral) labels. That way, we classify a sentence into \texttt{Positive} when the compound value $>$ 0.1, into \texttt{Neutral} when $-0.1 \leq compound\_value \leq 0.1$ , and into \texttt{Negative} when the compound value $<$ 0.1. In Figure \ref{subfig:sentimentComp} we can see a comparison of the number of positive, neutral, and negative reviews classified by each one of the libraries. As we can observe, all three libraries classified the majority of the reviews as \texttt{Positive} (2M, 2.02M and 1.95M), indicating also a skewed data. However, we see a more significant difference if we take a look at the \texttt{Neutral} and \texttt{Negative} reviews. On the one hand, regarding \textit{VADER}, we note that 245,538 reviews are \texttt{Neutral}, and 135,571 are \texttt{Negative}. On the other hand, \textit{TextBlob} classifies 371,348 reviews as \texttt{Neutral}, and 78,198 as \texttt{Negative}. Finally, \textit{Flair} library results show that 401,575 are \texttt{Negative} reviews, and only 516 reviews are \textit{Neutral}. Furthermore, in Figure \ref{subfig:boxplotSent} we can see a box plot comparing the sentiment values between the two libraries providing continuous values: \textit{VADER} and \textit{TextBlob}. As we observe, \textit{VADER} library produces higher sentiment values (median= 0.62, $Q1=0.45$, $Q3=0.77$) than TextBlob (median= 0.35, $Q1=0.14$, $Q3=0.6$).

\subsection{RQ3. Can we leverage NLP-driven topic analysis techniques to find themes that can be important for prospective learners?}

As described previously, we created two different topic finding models: the first one using words that describe the courses qualitatively; and the second one using words that described the courses' content. 
\subsubsection{Qualitative description model}

The first model that we built was the qualitative description one. After applying the LDA algorithm as described in Section \ref{sec:topicFin}, we obtained a set of coherence measures to determine which is the optimal number of topics for this concrete model. We determined 14 as the optimal number of topics, with a $C_v$ score of 0.20 and a $C_{umass}$ score of -3.88. A summary of each topic, including the five most important words related, can be found in Table \ref{tab:topicsModel1}.

\begin{longtable}{| p{.25\textwidth} | p{.75\textwidth} |} 
\hline
\textbf{Topic} & \textbf{Main terms} \\ \hline 
1 & Informative, easy, fun, enjoyable, helpful \\ \hline
2 & Personal, authentic, productive, write, child, book \\ \hline
3 & Basic, beginner, introduction, overview, simple \\ \hline
4 & Simple, easy, effective, clear, short \\ \hline
5 & Error, week, wrong, issue, outdated \\ \hline
6 & Voice, fast, slow, hard, monotone\\ \hline
7 & Money, worth, ad, free, quality \\ \hline
8 & Specialization, team, rigorous, introduction \\ \hline
9 & Step, explanation, detailed, easy, clear \\ \hline
10 & Exam, test, practice, certification, real \\ \hline
11 & Beginner, basic, advanced, intermediate \\ \hline
12 & Real, worth, life, apply, practical \\ \hline
13 & Theory, practical, theoretical, lab, hand \\ \hline
14 & Slide, powerpoint, visual, lack, repetitive \\ \hline
\caption{Summary of each detected topic regarding qualitative descriptions.}
\label{tab:topicsModel1}
\end{longtable}

\begin{figure}[!ht]
\includegraphics[width=\textwidth]{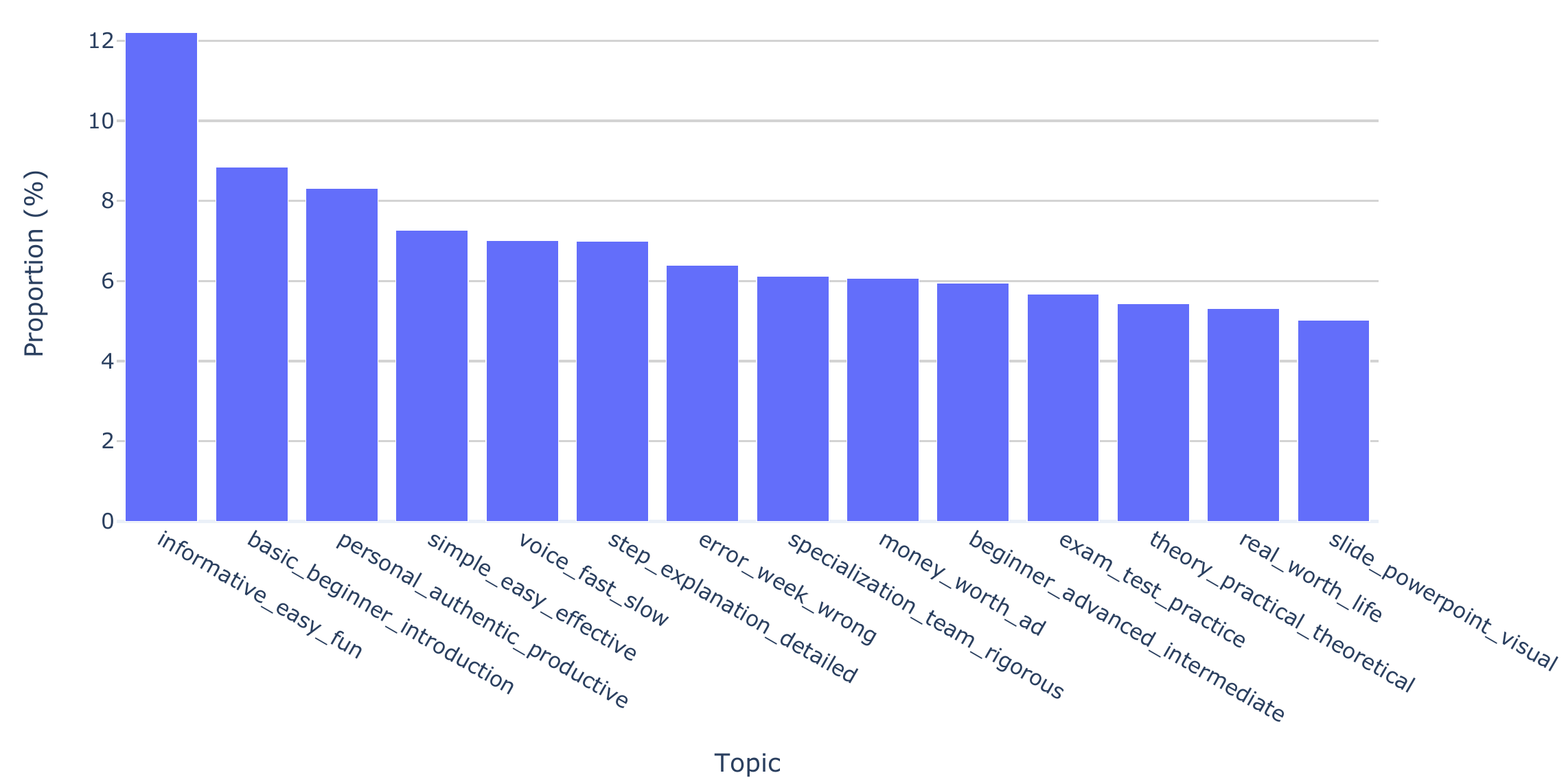}
\centering
\caption{Distribution of topics for the \texttt{QualitativeDescription} model.}
\label{fig:distrQual}
\end{figure}

We can see a wide variety of topics, such as the first one, described by the words ``informative,'' ``easy,'' or ``fun;'' or the second one, described by the words ``productive,'' ``personal,'' or ``authentic.'' Then, Figure \ref{fig:distrQual} shows the distribution of such topics across the entire collection of reviews. In the Figure, each topic is labeled by its three most important keywords (e.g., ``simple\_easy\_effective''). As we can observe, the most frequent topics are ``informative\_easy\_fun'' (12.2\%), ``basic\_beginner\_introduction'' (8.8\%) and ``personal\_authentic\_productive'' (8.3\%), and the less frequent topics are ``real\_worth\_life'' (5.3\%) and ``slide\_powerpoint\_visual'' (5.0\%).

\begin{figure}[!ht]
\includegraphics[width=\textwidth]{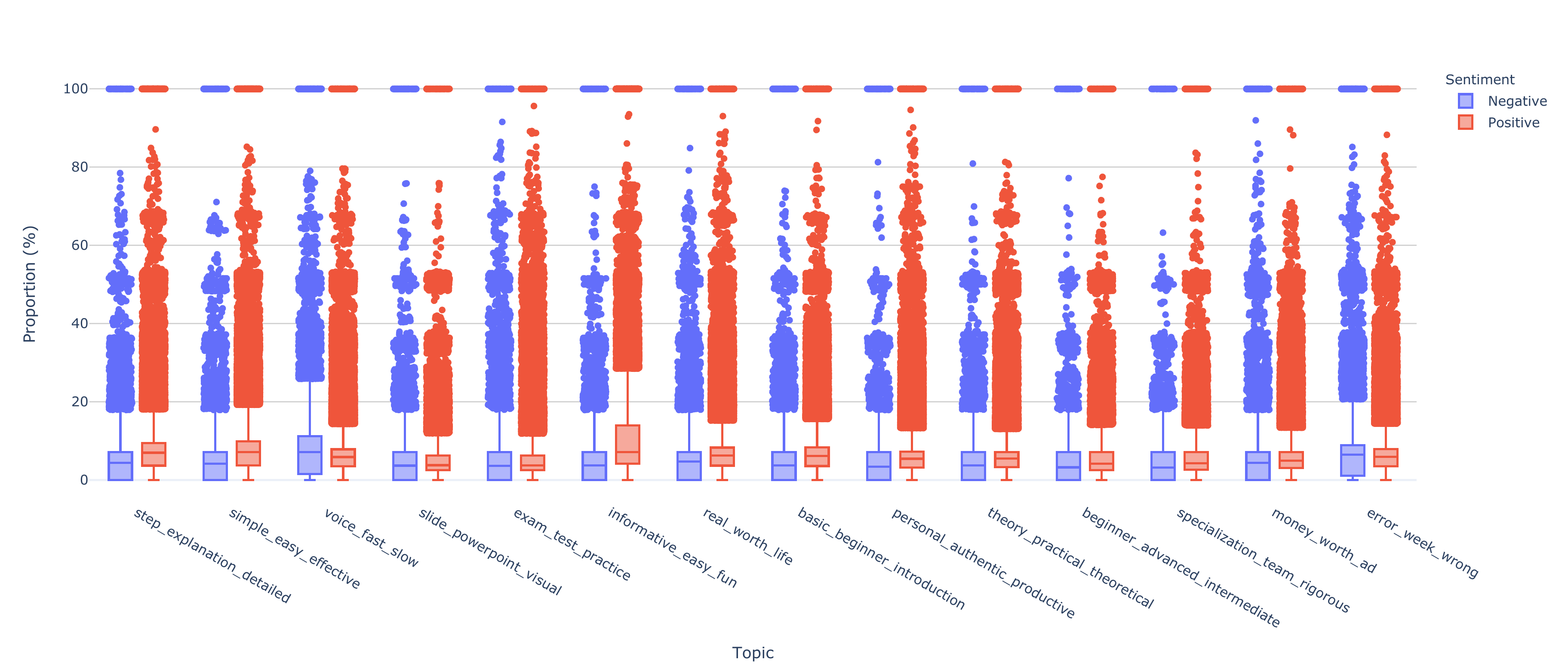}
\centering
\caption{Distribution of topics per course by sentiment values.}
\label{fig:sentimentQualit}
\end{figure}

To see the relationship between the qualitative topic model and the sentiment analysis, we calculated the distribution of topics in labeled courses based on their reviews' sentiment value. In this case, we used \textit{Flair} library, which, compared to other NLP packages, its sentiment classifier is based on a character-level LSTM neural network which takes sequences of letters and words into account when predicting. It is based on a corpus but in the meantime, it could also predict a sentiment for OOV (Out of Vocabulary) words including typos \citep{flairJust}. In Figure \ref{fig:sentimentQualit} we can see the distribution in both \texttt{Positive} and \texttt{Negative} labeled courses. As we can see, in the most frequent topics for positive courses, there are more positive topics such as ``informative\_easy\_fun'' (11.6\%) and ``simple\_easy\_effective'' (8.6\%). Then, in negative courses, the most frequent topics are the ones describing the course in a negative way, such as ``voice\_fast\_slow''(11.7\%) or ``error\_week\_wrong'' (10.4\%). However, since most of the topics are positive, courses with negative sentiment values also show a high frequency of positive topics.

Then, we conducted a MANOVA test including the 15 different topics, confirming that there is a significant difference in topics proportion when comparing  \texttt{Positive} and \texttt{Negative} courses ($F = 59.05$, $p < 0.0001$). Thus, we can see that there is a relation between the sentiment analysis and the topic modeling approach that we have performed in this research.

\subsubsection{Content model}

Moreover, the second model that we built was the one describing courses' content. After applying the LDA algorithm, we determined 14 as the optimal number of topics, with a $C_v$ score of 0.27 and a $C_{umass}$ score of -5.08. In this case, we have assigned a label to each topic based on the existing content categories in the corpus and the keywords of each topic. A summary of each topic, including its name, description, and the five most important words related, can be found in Table \ref{tab:topicsContent}.

\begin{longtable}{| p{.25\textwidth} | p{.4\textwidth} | p{.3\textwidth} |}
\hline
\textbf{Topic label} & \textbf{Description} & \textbf{Main terms} \\ \hline 
Health and lifestyle & Courses that contribute to physical, mental, and social well-being (balanced diet, getting more rest, doing physical exercise...). & Life, exercise, food, calm, meditation \\ \hline
Programming & Courses that aim to teach programming knowledge. & Exercise, programming, code, language, programmer \\ \hline
Machine and deep learning & Includes courses that teach machine and deep learning techniques. Both aim the Artificial Intelligence (AI) to learn from data and then apply what they have learned to make informed decisions. & Machine, deep, learning, neural, network \\ \hline
Cloud computing & Courses teaching how to use cloud computing services (e.g., Microsoft Azure, Amazon Web Services). & Feature, topic, angular, azure, api \\ \hline
Investing \& Trading & Includes courses that aim to teach investing and trading knowledge and techniques. Although both investing and trading are methods of attempting to profit in the financial markets, investing often takes long-term approach and trading usually takes a short-term approach. & Business, trading, market, trade, financial \\ \hline
Music & Courses related with the music field, such as music production, how to sing, or how to play any instrument. & Music, play, audio, song, piano \\ \hline
Network \& Security & Courses that aim to teach security and networks related content, such as how to prevent a hacking attack or how to provide more security to your devices. & Security, software, hack, project, management \\ \hline
Language learning & Courses dedicated to learn any language (e.g., Spanish, French). & Language, speak, accent, pronunciation, chinese \\ \hline
Finance \& Accounting & Content related with accounting, which is an essential tool for providing information for decision-making, as well as for the evaluation of decisions previously made, and finance, that must seek resources at a reasonable cost and use them efficiently \citep{mele2017ethics}. & Management, business, leadership, law, economic \\ \hline
Arts \& Crafts & Includes courses teaching knowledge about decorative design and handicraft. & Draw, art, paint, artist, oil \\ \hline
General health & Includes different general health sub-topics such as economic aspects, privacy, philosophy, or cultural aspects. & Life, topic, business, psychology, philosophy \\ \hline
Data science & Courses aiming to teach data science, which have emerged as a new and important discipline, and it can be viewed as an amalgamation of classical disciplines like statistics, data mining, databases, and distributed systems \citep{van2016data}. & Datum, science, statistical, excel, visualization \\ \hline
3D \& Animation & These courses include 3D modeling and animation approaches, modeling objects or characters and utilizing motion in order to bring those characters, objects and more to life. & Design, software, graphic, animation, software \\ \hline
Game development & Includes different approaches that are part of developing a video game. & unity, game, unreal, software, engine \\ \hline
\caption{Summary of each detected topic regarding courses' content.}
\label{tab:topicsContent}
\end{longtable}

\begin{figure}[!ht]
\includegraphics[width=\textwidth]{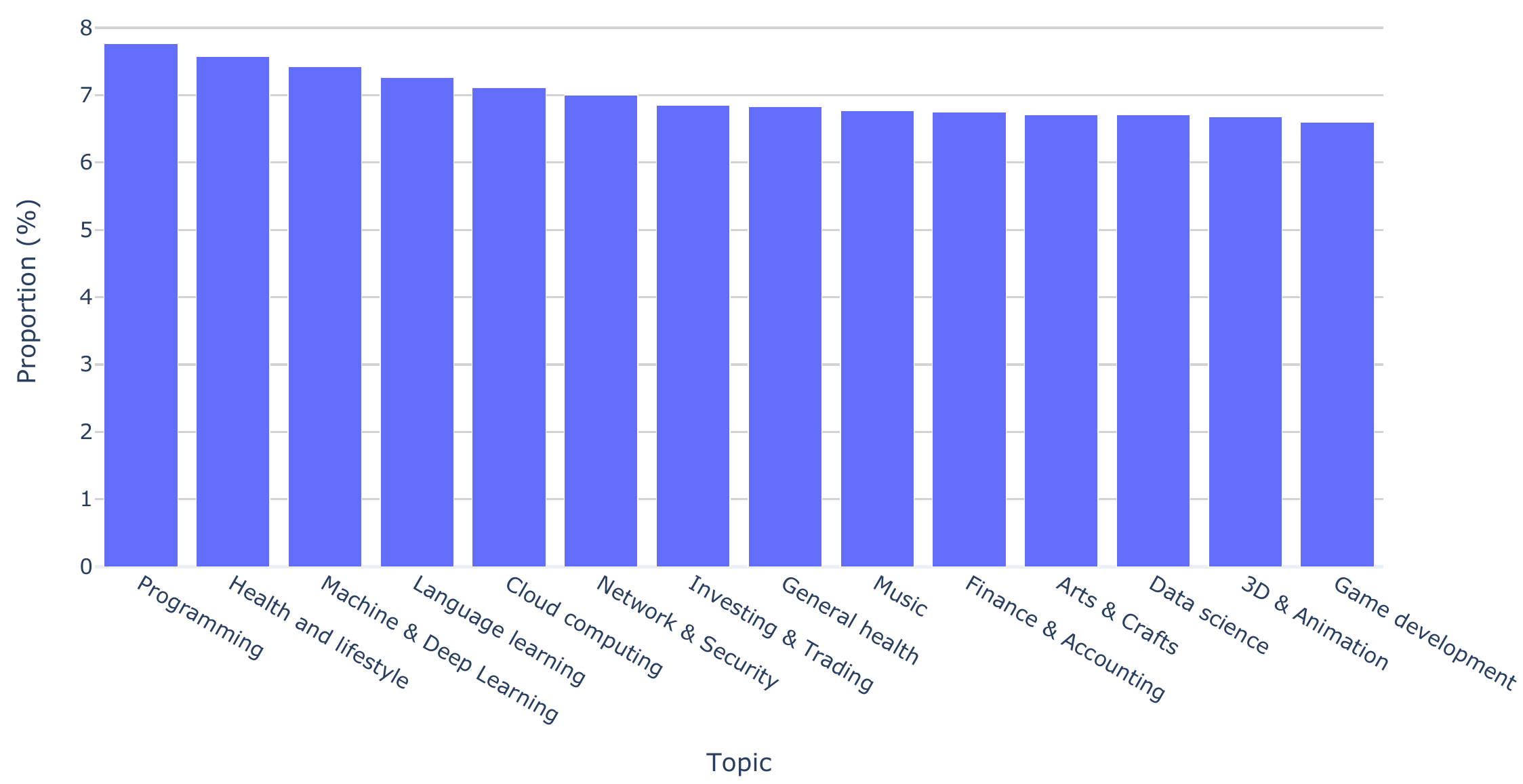}
\centering
\caption{Distribution of topics for the \texttt{Content} model.}
\label{fig:distrCont}
\end{figure}

Again, we see a large variety of topics referring to different content areas that courses aim to teach, such as ``Programming,'' ``Music,'' or ``Language learning.'' Then, Figure \ref{fig:distrCont} shows the distribution of such topics across the entire collection of reviews. In this case, we see that there is a very uniform distribution of topics, since all of them have a similar value. Although there is a wide variety of knowledge areas among the different courses, there is only a 1.5\% of difference between the most and less frequent topics.

\subsection{RQ4. Can we use these models to effectively characterize MOOCs based on the open reviews?}
\label{subsec:RQ4}

To evaluate if we can effectively characterize MOOCs using the models developed in this research, we chose a set of four MOOCs belonging to different topics in order to analyze them:

\begin{itemize}
    \item \textbf{Finance for Non-Finance Professionals}. The original category of this course is ``Business'' and has 342 available reviews.
    \item \textbf{Using Python to Access Web Data}. The original category of this course is ``Computer science'' and has 5,748 available reviews.
    \item \textbf{Mind Control: Managing Your Mental Health During COVID-19}. The original category of this course is ``Health'' and has 1,183 available reviews.
    \item \textbf{Teach English Now! Foundational Principles}. The original category of this course is ``Social sciences'' and has 2,447 available reviews.
\end{itemize}

\begin{figure}[!ht]
\includegraphics[width=\textwidth]{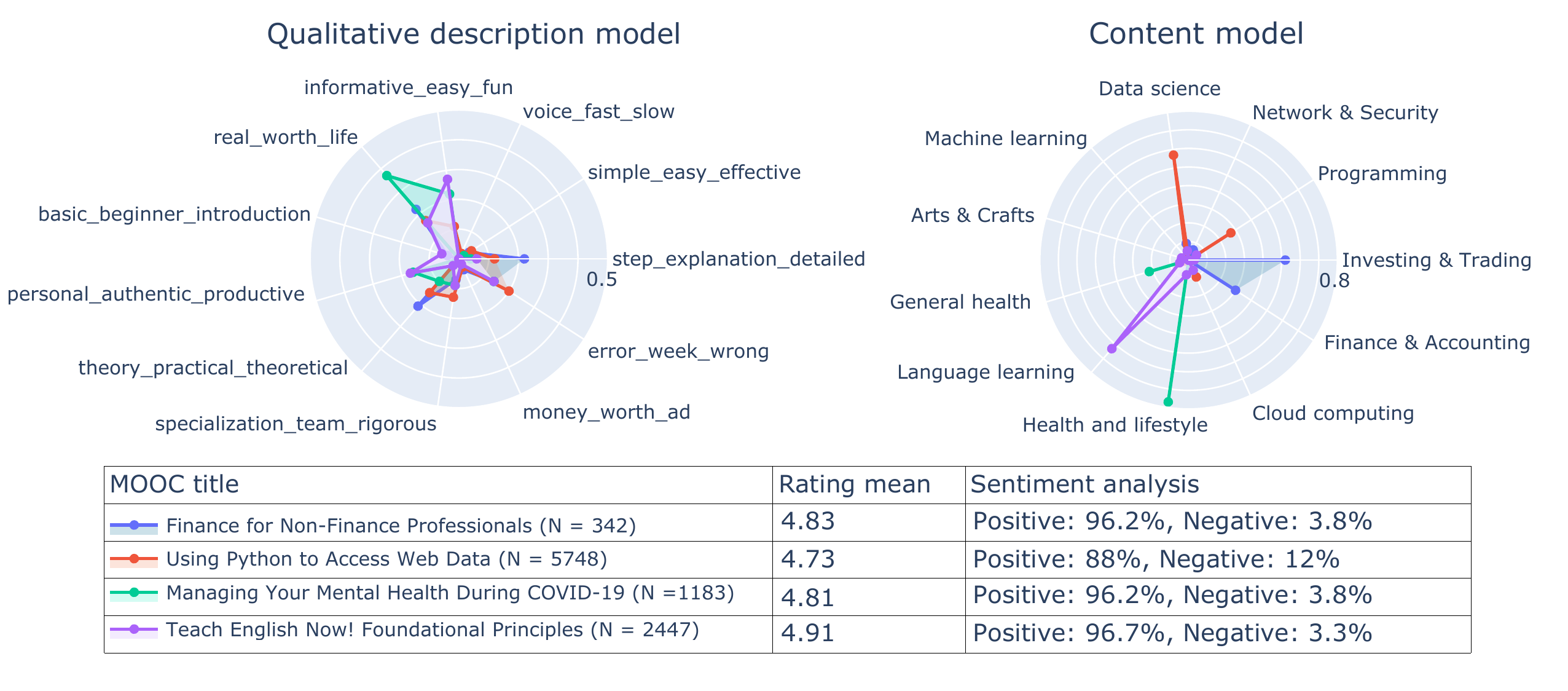}
\centering
\caption{Rating mean, sentiment and topic analysis of four different MOOCs.}
\label{fig:caseStudy}
\end{figure}

As we can see in Figure \ref{fig:caseStudy}, all four courses have a rating mean above 4.7 stars, which confirms the biased data that we previously highlighted. Furthermore, we also see a positive correlation between the rating mean and the sentiment analysis results of the courses (calculated using \textit{Flair} library), since the courses showing a higher rating mean also show a higher percentage of \texttt{Positive} labeled reviews. For example, the business related course has a rating mean of 4.83/5, and the 96.2\% of its reviews are labeled as \texttt{Positive}, meanwhile the Python related course has a rating mean of 4.73/5, and the 88\% of its reviews are labeled as \texttt{Positive}.

Then, we evaluated the two topic models obtained through NLP on the open reviews. For the qualitative description model on the left, the COVID-19 and English courses have a high proportion of the topics ``real\_worth\_life,'' ``informative\_easy\_fun,'' and ``personal\_authentic\_productive,'' meanwhile the Python related course has a higher proportion of the topic ``error\_week\_wrong,'' which indicates that reviews are revealing more problems in this last course than in the others. Furthermore, this finding is also confirmed by the fact that this course is the one with the lowest rating mean and sentiment analysis result of the selected courses. In addition, the topics that have a high proportion in the COVID-19 and English courses reveal additional positive aspects of these courses, which is the personalized and real application that they show. Regarding the content model, we see that our approach obtains an accurate result when modeling the content of each course automatically. For example, we see that the health related course, which was originally classified within the ``Health'' category, has a very high proportion of the topics ``General health'' and ``Health and lifestyle.'' Furthermore, the business related course, which was originally labeled as a ``Business'' course, has a very high proportion of the topics ``Investing \& Trading'' and ``Finance \& Accounting.'' These results validate that our models can automatically add additional and reliable information for learners to select the most appropriate course to their preferences.

\section{Discussion}
\label{sec:discussion}

\subsection{Interpretation of results}

This study has four research objectives. First, the study aims to analyze the numeric ratings in the collection in order to see if they provide any valuable information to learners. Second, the study aims to perform NLP-driven sentiment analysis on textual reviews to see if this analysis provides any valuable information to learners, comparing it with the previous numeric ratings. Then, the study aims to perform topic analysis on textual reviews to find the most representative themes across the corpus and see if these themes provide any valuable information to learners. To achieve these research objectives, this study used around 2.4 million reviews from learners who participated in 93,678 different courses across four MOOC platforms. Major findings revealed in this study are reported in this section and discussed in conjunction with the existing MOOC literature. Finally, this research performed an analysis trying to characterize four different MOOCs, applying the models developed in this study.

In Section \ref{sec:relWork} we identified some previous studies that also tried to analyze existing reviews in the MOOCs field. For example, \cite{chen2020moocs} analyzed 1920 reviews from 339 computer science courses, finding that 64.2\% of the reviews in their data collection were 5-star reviews, while 16.7\% were 4-star reviews. In addition, \cite{gamage2016star} analyzed 137 courses from Coursera, edX, FutureLearn, Udacity, Iversity, Canvas, and few Independent MOOCs platforms from Universities. They found that 18.24\% of the courses had a 5-star rating, while 43.8\% of the courses had a 4-star rating. The results of our research also confirm that the majority of reviews are 5-star and 4-star ratings. This bias is evident in the fact that most courses receive a majority of 5-star ratings. Further, using the 5-star rating system may provide limited insights in separating good from great products or supplies, and star ratings may be less meaningful to differentiate what learners like and dislike about a course design \citep{li2021key}. Instead, this research also explores sentiment analysis and topic modeling to gather insights into different course elements that were positive or negative to the learner experience.

The majority of MOOC reviews are less than 50 words long, and such short text is likely to create poor context-dependency, insufficient co-occurrence information, and sparse feature matrix, and result in an awful performance of LDA \citep{qi2021evaluating}. \cite{knoos2021sentiment} analyzed 28,000 reviews scraped from five courses within the fields of data science, and found that the average length of the extracted reviews was 103 characters. Many reviews consisted of only one or two adjectives such as ``good'' or ``nice.'' In addition to having the biggest dataset analyzed in the area, our research found that the average length of the extracted reviews was 134.6 characters, which is larger than the typical average length of a review. In addition, since the coherence score is only an indicator of the model performance, we made iterations and manual reviews of the topics to ensure the relevance and coherence of the topics. 

Our sentiment analysis results confirm that we have a skewed data, revealing also a positive correlation with the numeric ratings. Previous studies \citep{knoos2021sentiment} used \textit{VADER} sentiment analysis to predict the sentiment of reviews, finding that the library could predict the sentiment of positive reviews with a 93\% accuracy. However, the accuracy in the prediction of negative and neutral reviews was less than 60\%. Thus, since there might be some negative reviews classified as positive, we can only estimate the factors and topics that learners had a negative sentiment towards. 

\cite{chen2020moocs} performed structural topic modeling and identified some interesting topics such as ``course level.'' ``teaching style'' or ``problem solving,'' meanwhile \cite{deng2021key} identified ``Learning'' as the most salient theme. Moreover, a closer inspection on that concrete theme showed that learners placed high importance on making connections between MOOC content and their everyday lives. In our research, we found that reviews also made emphasis on the importance of MOOCs application in real life, with frequent topics and important keywords, such as topic 9 (``practice,'' ``real'') and topic 12 (``real,'' ``worth,'' ``life,'' ``practical''). Seeing this trend, we recommend that MOOCs content should be focused on a more practical approach, being applied in real life contexts and environments, and instructional conditions in MOOCs should be engineered to facilitate the acquisition of knowledge that can be transferred to real-world practices.

In their study, \cite{chen2020moocs} explored the relationship between the ratings and the topics identified, finding that negative reviews tended to relate more to issues such as course assessment, learning tools and platforms, and critique, while positive reviews concerned more about issues such as course levels, course organization, and learning perception. In our research, since the correlation identified between sentiment analysis results and ratings is a moderate one, we also decided to explore the relationship between sentiment analysis and our topic modeling approach, finding that courses with positive sentiment value tend to be related to positive topics, while courses with negative sentiment value tend to be more related with negative topics.

Furthermore, we can take a look at both of our topic models and discuss the coherence and usability of the topics revealed. Regarding the qualitative description model, we see that some topics could be really useful for students to identify the courses' key features, such as topic 14 (``slide,'' ``powerpoint,'' ``lack,'' ``repetitive'') identifying courses that might be not as well designed as other, or topic 2 (``personal,'' ``authentic,'' productive'') highlighting the personalization that many students missed in other courses \citep{gutl2014attrition}. However, we also identify other topics that might not add useful information, such as topic 11 (``beginner,'' ``intermediate,'' ``advanced'') showing keywords that are confusing. Another example is topic 13, mixing the keywords ``theory'' and ``practical,'' which also do not clarify any information. Regarding the content model, we believe that the topics revealed by the model are appropriate in comparison with the original categories defined by the courses themselves, and also consistent between them. However, a topic model with a larger number of topics may discover new ones that are not included in our current analysis.

\subsection{Implications and limitations}
This research's findings validate that the existing open MOOC review systems contains courses with extremely positive ratings, making the decision-making process challenging. A potential MOOC student will face thousands of highly-rated MOOCs for popular topics such as Python or Excel, wherein the majority of courses ratings are around 4.7 stars or above, exacerbating what could be described as a ``choice paradox'' \citep{scheibehenne2010can}. A challenge that no MOOC provider has taken seriously is that all review and search systems are far from vertical-specific and have not improved significantly since MOOCs became more popular in 2012. This is not surprising, as providers have no incentives to improve the decision-making process and mainly focus on widening their offer and reach.

It is unknown whether providers mostly display courses with the highest rating by curation or if a structural decision pushes the overall distribution of the system to a standard that does not reflect reality. Moreover, there is a clear economic incentive for MOOC providers to only display courses in the high four-stars range. A recent study from McKinsey found that even the slightest improvements in the star rating score (+0.2) in the retail sector could trigger a 37\% increase in sales throughout a product life cycle \citep{Mckinsey2020}. The bias mentioned above, combined with an online education system with completion rates as low as 7\%, creates the perfect scenario for MOOC students to be skeptical about the quality promoted by these providers. Ultimately eroding trust in the system and leading to inefficient use of time and money throughout the online educational journey. 

This study also has some limitations. Non-English reviews were excluded from this study. An analysis of non-English reviews may provide additional insight into the learning experience idiosyncratic to these learners (e.g., individuals who use non-English MOOC platforms, individuals whose primary language is not English) \citep{deng2021key}. Future research could overcome this limitation by analyzing non-English reviews to provide a more comprehensive and complete understanding of the learning experience in MOOCs. Moreover, the reviews considered for the analysis were not further filtered by length and contextual richness. Arguably, further research could utilize a subset of reviews that would provide more context, and ergo the results of a new topic labeling system could provide more nuance for the student.

There are reasons to believe artificial intelligence and community-driven tools can help improve the overall system. In this vein, the approach in our work, using topic modeling, and the standardization and decentralized review systems can provide more transparency to the decision-making process. Moreover, we can foresee increasing interest in platforms leveraging the aforementioned techniques and want to help students in their educational journey. Thus, creating a better system that increases the Return of Education (ROE).  Last but not least, it is essential to reckon that the online education revolution will be only as good and transformative as the providers want it to be. However, there is also agency on the students poised to seek more transparency in online education systems, which are only becoming more popular as we enter the post-pandemic era.

\section{Conclusions}
\label{sec:conclussion}

In this study, we aimed to answer four research questions in order to support learners' course selection in MOOCs, using the largest dataset so far in the literature. First, we analyzed the numeric ratings in our data collection, finding a biased dataset with a majority of 5-star and 4-star ratings. Then, we performed sentiment analysis on textual reviews using three pre-trained models from different libraries. Results suggests that there is a positive correlation between the compound sentiment values and the numeric ratings. All three libraries show that the majority of textual reviews (with results varying between 80.8\% and 83.8\%) are classified as positive reviews. Then, we built two models based on a topic modeling approach. The first model used words that described the courses in a qualitative way, and revealed some interesting topics and keywords such as ``informative\_easy\_fun,'' ``money\_worth\_ad'' or ``slide\_powerpoint\_visual.'' This approach can effectively characterize courses, thus helping learners to decide which course to choose, highlighting features that have been important for learners over time. The second model used words that described the courses' content, and revealed frequent topics such as ``Health and lifestyle,'' ``Programming,'' or ``Cloud computing.'' Since each course can cover different areas, this approach can help learners to discover some hidden topics in existing courses, based on other learners' reviews. We also calculated the distribution of each topic, which allowed us to see which were the most frequent and the less frequent topics across the entire collection. Moreover, we explored the relationship between the sentiment analysis results and the qualitative topic modeling approach, finding that courses with positive sentiment value tend to be related to positive topics such as ``informative\_easy\_fun'' or ``real\_worth\_life,'' while courses with negative sentiment value tend to be more related with negative topics like ``error\_week\_wrong.'' Finally, we tried to characterize four different MOOCs using our models, which confirmed the positive correlation between the sentiment analysis and the numeric rating, but also the difference in topic characterization when comparing courses with higher and lower sentiment analysis results.

As part of our future work, we would like to validate the \texttt{Content} topic model against the original content category in every course to see the accuracy of the model. Moreover, since the typical limited size in reviews could be a problem for the LDA model, future studies could consider only large reviews in order to explore more relevant or meaningful topics. In addition, some of the pre-trained sentiment models tend to fail in some classifications (specially referring to neutral and negative reviews). Thus, in future studies, the design of a lexicon that better takes into account negative words could improve the analysis and results obtained. Finally, another step would be the development of a real recommendation system to validate the appropriateness of these discovered topics for its real purpose: allow students to identify the best courses among the almost infinite variety of them. To wrap up, we believe that there is an opportunity to leverage current approaches and create more powerful, yet simpler reviewing systems, using techniques similar to the ones applied in this research. Thus, being able to really identify the best courses out there, and turn a blind eye to the current highly biased MOOC review systems.

\bibliographystyle{elsarticle-harv} 
\bibliography{bibliography}





\end{document}